\documentclass[epj]{webofc}
\usepackage[utf8]{inputenc}
\usepackage[varg]{txfonts}   
\usepackage{booktabs}
\usepackage{xcolor}
\definecolor{darkred}{rgb}{0.4,0.0,0.0}
\definecolor{darkgreen}{rgb}{0.0,0.4,0.0}
\definecolor{darkblue}{rgb}{0.0,0.0,0.4}
\usepackage[bookmarks,linktocpage,colorlinks,
    linkcolor = darkred,
    urlcolor  = darkblue,
    citecolor = darkgreen]{hyperref}
\usepackage{subfigure}
\wocname{EPJ Web of Conferences}
\woctitle{Lattice2017}
\newcommand{\be}{\begin{equation}}
\newcommand{\ee}{\end{equation}}

\begin{document}
%
\selectlanguage{english}
\title{%
Temperature-dependence of the QCD topological susceptibility
}
\author{%
\firstname{Tamas G.}  \lastname{Kovacs}\inst{1}\fnsep\thanks{Speaker,
  \email{kgt@atomki.mta.hu}. Acknowledges financial support under contract
  number OTKA-K-113034. Thanks Z.\ Fodor, S.D.\ Katz and K.K.\ Szabó for
  numerous helpful discussions during the preparation of this talk. Most of
  the computations presented here were performed on JUQUEEN at
  Forschungszentrum Jülich. }
}
\institute{%
Institute for Nuclear Research of the Hungarian Academy of Sciences, H-4026
Debrecen, Hungary
}
\abstract{%
  We recently obtained an estimate of the axion mass based on the hypothesis
  that axions make up most of the dark matter in the universe. A key
  ingredient for this calculation was the temperature-dependence of the
  topological susceptibility of full QCD. Here we summarize the calculation of
  the susceptibility in a range of temperatures from well below the finite
  temperature cross-over to around 2 GeV.  The two main difficulties of the
  calculation are the unexpectedly slow convergence of the susceptibility to
  its continuum limit and the poor sampling of nonzero topological sectors at
  high temperature.  We discuss how these problems can be solved by two new
  techniques, the first one with reweighting using the quark zero modes and
  the second one with the integration method.
}
\maketitle
\section{Introduction}\label{intro}

One of the most viable candidates for dark matter is the axion
\cite{Moore:2017ond}. However, experimental search for this so far
hypothetical particle is seriously hampered by the lack of information
regarding its mass. The axion would also solve the long-standing strong CP
problem of QCD due to its coupling to the QCD topological charge. A crucial
ingredient for estimating the axion mass is the temperature-dependence of the
QCD topological susceptibility, a quantity that proved to be notoriously hard
to compute.

There are two main difficulties involved in the computation of the topological
susceptibility at high temperature. Firstly, there are unexpectedly large
cut-off effects that become worse as the temperature goes up. The origin of
this problem can be easily understood from the underlying physics. At high
temperature typical instantons become small, as their size is controlled by
$1/T$, the temporal extent of the system. Small instantons are badly resolved
by a non-chiral Dirac operator and the would-be topological zero modes are far
away from zero. As a result, the quark determinant cannot provide the
appropriate suppression of higher topological sectors. As we will see, in a
simulation with conventional methods, infeasibly fine lattices would be needed
to resolve small instantons.

The second difficulty of the computation stems from the fact that above the
finite temperature cross-over to the quark-gluon plasma, the topological
susceptibility falls sharply with increasing temperature. The resulting tiny
susceptibility is hard to measure since even in large volumes the system
rarely visits non-trivial topological sectors. We emphasize that this is a
physical property of the system and has nothing to do with the autocorrelation
time of the used algorithm. However, there is also a connected algorithmic
problem that makes the situation even worse, and this is the slow topology
change on fine lattices. These factors together make it prohibitively
expensive to gather enough statistics for the susceptibility with current
resources.

In the present work, based on Ref.\ \cite{Borsanyi:2016ksw}, we compute the
QCD topological susceptibility starting from the $T=0$ region through the
transition temperature around $150$~MeV \cite{Borsanyi:2010bp}, up to $2$~GeV,
that is the whole temperature range relevant for axion cosmology. We offer
solutions to both problems hampering the calculation of the susceptibility. We
show that even without resorting to extremely fine lattices, the incorrect
suppression of higher topological sectors can be cured by a suitable
modification of the quark determinant. This involves shifting would-be zero
modes to exactly zero and reweighting configurations with the ratio of the
original determinant and the one with shifted zero modes.

The second problem, that of insufficient statistics for higher topological
sectors, can be solved by noting that derivatives of the susceptibility (with
respect to temperature and quark mass) are much easier to calculate than the
susceptibility itself. Indeed, these derivatives can be written in terms of
the average action and the quark condensates measured in different topological
sectors. Using these derivatives, the susceptibility can be integrated up to
any point in parameter space, starting from a point (low temperature, heavy
quarks) where a direct calculation is feasible. This procedure results in the
so called integral method that was independently also suggested for the
quenched case by Frison et al. \cite{Frison:2016vuc}.

In the following, using reweighting and the integral method, we present a
calculation of the QCD topological susceptibility in the temperature range $0
< T < 2$~GeV. Our final results pertain to the real physical situation
including the continuum limit and the effects of dynamical $u,d,s$ and $c$
quarks with their physical masses along with a correction for the $u-d$
isospin splitting.

\section{Unusually large cutoff effects}\label{sec-1}

For the simulations that we present here we used the tree-level improved
Symanzik gauge action together with staggered fermions on four times
stout-smeared gauge links \cite{Morningstar:2003gk}. The continuum limit was
based on simulations with temporal lattice sizes of $N_t=8,10,12,16$ and
$20$. For the lowest temperature range, up to $T=250$~MeV our simulations
include $2+1$ quark flavors with physical masses and for higher temperatures
where it contributes significantly, we also included the charm quark. Some of
our simulations were done with overlap quarks \cite{Neuberger:1997fp}. As
explained later in the text, at intermediate stages some of our calculations
involved simulations with heavier than physical $u,d$ quarks up to the strange
quark mass $(N_f=3)$, but this was only needed for the integral method, no
extrapolation in the quark mass was involved. For more details of our setup we
refer the reader to \cite{Borsanyi:2016ksw,Borsanyi:2015zva} and
\cite{Borsanyi:2012xf}.

In this section we show why the topological susceptibility at high temperature
suffers from unusually large cutoff effects.  We will also describe how to
solve this problem without utilizing extremely fine lattices that would make
it prohibitively expensive to obtain the continuum limit with present-day
resources.

\subsection{The problem}

\begin{figure}[thb]
  \centering
  \includegraphics[width=0.8\textwidth,clip]{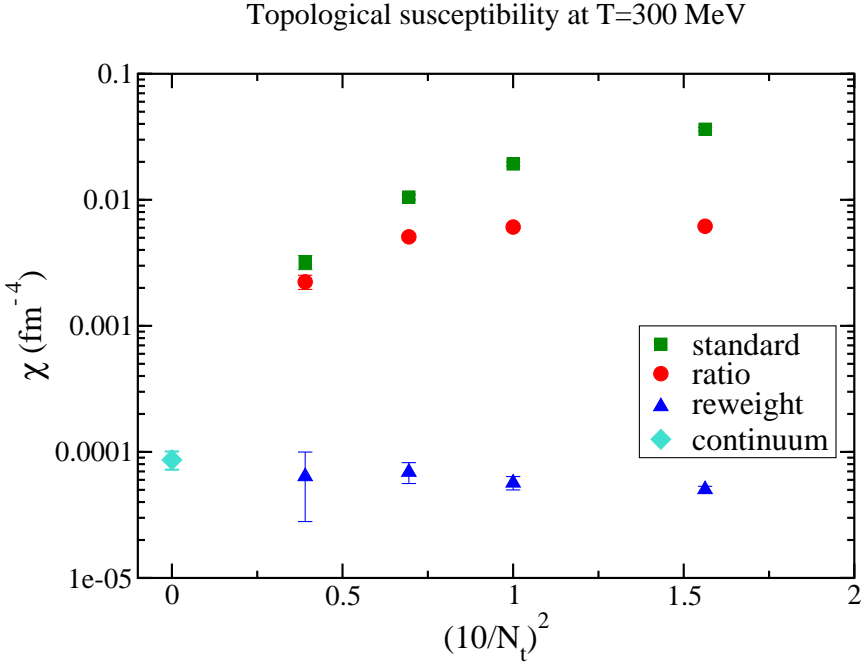}
  \caption{The dependence of the topological susceptibility on the lattice
    spacing at $T=300$~MeV, obtained with two conventional methods (see text)
    and reweighting (to be discussed later). The horizontal axis is
    proportional to the lattice spacing squared. The continuum extrapolated
    point at zero was obtained using our reweighting method.}
  \label{fig:T300_rew}
\end{figure}

To illustrate the cutoff effects we face here, in Fig.~\ref{fig:T300_rew} we
show the topological susceptibility at $T=300$~MeV, measured using
conventional methods on lattices of different coarseness corresponding to
$N_t=8,10,12$ and 16. This is a typical setup for taking the continuum limit
in present-day QCD thermodynamics calculations. The data labeled ``ratio'' was
obtained by normalizing the raw data with the zero temperature
susceptibility at the same lattice spacing \cite{Bonati:2015vqz}. The first
three points (from the right) of this data set appear to be nicely scaling,
but the point on the finest lattice reveals that scaling will probably set in
only for much finer lattices. Indeed, for comparison we also included the
correct continuum extrapolated value that our reweighting method (to be
described) yields. Clearly, nothing comparable could have been obtained by
extrapolating the {\it standard}  or {\it ratio} data sets in the figure.

What is the the source of these unexpectedly large scaling violations for the
susceptibility at high temperature? In particular why does one need much finer
lattices at high temperature than at zero temperature for the continuum limit
of the susceptibility? The answer to this question can be given by examining
the typical size of instantons contributing to the susceptibility at different
temperatures. At zero temperature the instanton size is limited by
non-perturbative QCD physics and as a result typical instantons have a radius
around 0.3-0.4~fm \cite{Hasenfratz:1998qk}. In contrast, above the cross-over
to the quark-gluon plasma, the temporal size of the system becomes too small
to accommodate these instantons. Therefore, at high temperature the typical
instanton size is controlled by the temporal size of the system, $1/T$ and the
size of typical instantons dominating the susceptibility becomes smaller,
proportionally to the inverse temperature.

In QCD with light dynamical quarks an important factor influencing the
topological susceptibility is the suppression of higher topological sectors by
the quark determinant. This suppression can mostly be attributed to the zero
eigenvalues (zero modes) of the quark Dirac operator in the presence of
topological charge. In the topological sector with charge $Q$ each quark
flavor has (at least) $|Q|$ zero eigenvalues \cite{Atiyah:1968mp} contributing
a factor $m_f^{|Q|}$ to the determinant, where $m_f$ is the mass of the given
quark flavor. However, this is true only in the continuum limit. Topological
zero modes of a non-chiral lattice Dirac operator do not occur at exactly
zero. For a given lattice Dirac operator, how far an instanton zero mode is
shifted away from zero depends mostly on the size of the corresponding
instanton; the larger the instanton is in lattice units, the better it can be
resolved by the lattice Dirac operator, the closer the would be zero mode is
to zero.

Putting this all together we can understand the origin of large scaling
violations. At high temperature typical instantons are small and the
corresponding zero modes $\lambda_0$ are farther away from zero. As a result,
the suppression factor they contribute in the quark determinant,
$\lambda_0+m_f$ can be significantly different from the continuum factor
$m_f$ and higher topological sectors are not sufficiently suppressed. 

For a more quantitative understanding let us assume that we keep $N_t$, the
temporal extent of the system in lattice units fixed and set the temperature
by adjusting the gauge coupling $\beta$. Moving along the line of constant
physics the quark mass in physical units remains constant (up to logarithmic
corrections). However, the instanton zero modes $\lambda_0$ (again in physical
units) will increase proportionally to the temperature. This is because the
resolution of typical instantons do not change, their size remains roughly
$N_t$ in lattice units and then the zero modes in lattice units $\lambda_0 a$
also do not change which in turn implies that 
\be
  \lambda_0 \propto \frac{1}{a} \propto T
\ee
This means that in this setup at higher temperatures the zero mode factors
$\lambda_0 + m_f$ are more and more dominated by the increasing $\lambda_0$
and as a result, higher topological sectors are not properly suppressed. The
foregoing discussion also suggests that if we wanted to compensate this effect
by working on finer lattices, the lattice spacing would need to be made
smaller proportionally to the inverse temperature. In the temperature range at
hand this is clearly not feasible with presently available resources. It is
also clear that if instead of increasing $\beta$, the temperature is raised by
choosing smaller values of $N_t$, the resolution of typical instantons and, as
a result, the scaling violations will become even worse.

\subsection{Reweighting}

Understanding the problem in this way also helps to find the solution. The
appropriate suppression of higher topological sectors can be restored by
shifting the would be zero modes to zero, where they would occur in the
continuum limit. 

As a first step we have to identify the would be zero modes produced by the
topological charge. To this end we determine the topological charge of each
configuration using the Wilson flow \cite{Luscher:2010iy} and extracting the
charge $Q$ from the smoothed gauge field. Next we determine the $4|Q|$
smallest magnitude eigenvalues of the staggered Dirac operator. We identify
these as the would-be zero modes and shift them in the quark determinant to
zero. This amounts to a reweighting of the given configuration with the factor
\be
 w[U] = \prod_f \prod_{n=1}^{4|Q|} 
         \left( \frac{m_f}{\lambda_n[U] + m_f}\right)^{1/4},
\ee
where the product runs over the light quark flavors $(u,d,s)$. Notice that in
a gauge field background with topological charge $Q$ the continuum Dirac
operator would have $|Q|$ exact zero modes. To take into account the expected
degeneracy of staggered quarks in the continuum limit, here we use the $4|Q|$
smallest magnitude modes for the reweighting, but, as usual in staggered
simulations, also take the fourth root to restore the proper number of flavors
in the determinant.

\begin{figure}[thb]
  \centering
  \includegraphics[width=0.8\textwidth,clip]{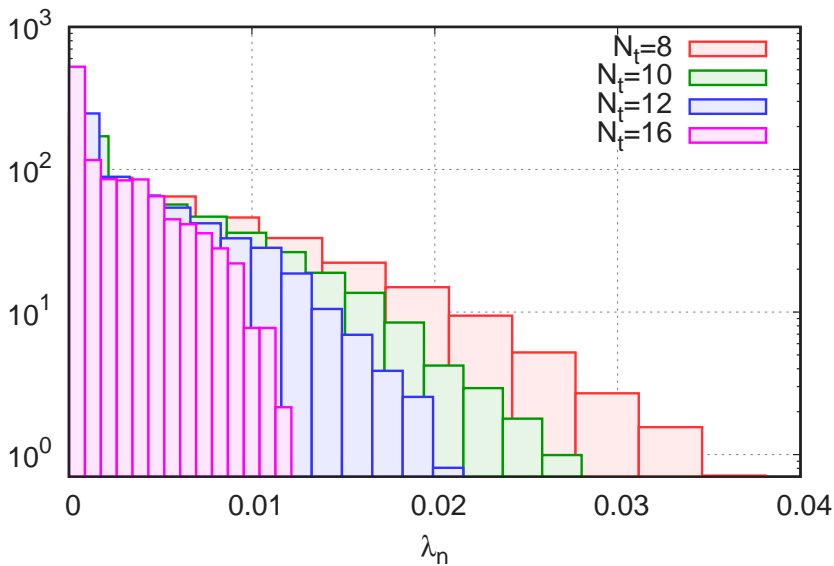}
  \caption{The distribution of the would-be zero modes of the staggered Dirac
    operator for different lattice spacings.}
  \label{fig:rew}
\end{figure}

\begin{figure}[thb]
  \centering
  \includegraphics[width=0.8\textwidth,clip]{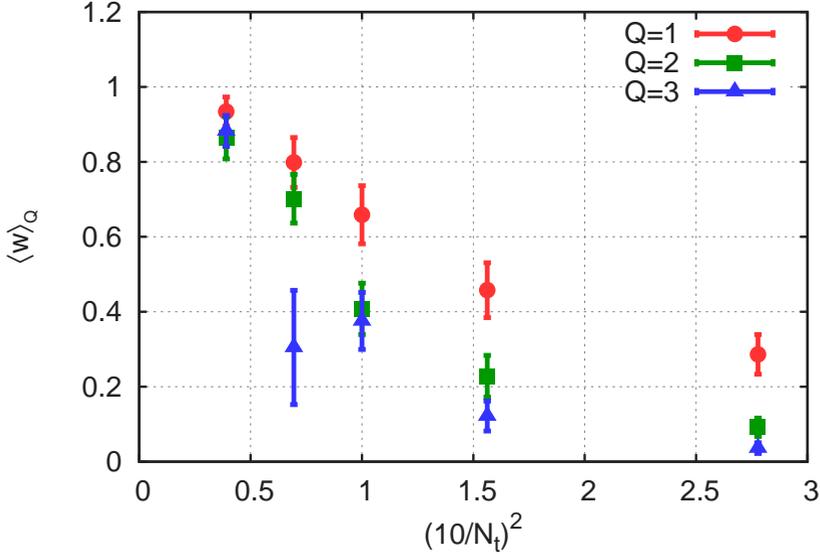}
  \caption{The average reweighting factors for different charge sectors as a
    function of the lattice spacing squared for the $N_f=3$ system at
    $T=300$~MeV.}
  \label{fig:wq1}
\end{figure}

The reweighting is a non-local modification of the quark action, however, in
the continuum limit the reweighting factors tend to unity and the original
quark action is restored. This happens because on finer lattices the
topological would be zero modes get closer to zero. We demonstrate this in
Fig.~\ref{fig:rew} by plotting the distribution of the zero modes on lattices
with different lattice spacings. The distribution can be seen to become
narrower, typical lattice zero modes move closer to zero as the lattice
becomes finer. As demonstrated in Fig.~\ref{fig:wq1}, we also explicitly
checked that the reweighting factors tend to unity as the continuum limit is
taken.

The reweighted susceptibility scales much
better, as is apparent in Fig.~\ref{fig:T300_rew}. This indicates that the
lack of scaling of the original non-reweighted data is indeed due to the
failure of the Dirac operator to capture the zero modes properly and thereby
failure to provide the proper suppression of higher topological sectors. 

\section{The integral method}

As we already mentioned, the second challenge in computing the topological
susceptibility at high temperature is the bad sampling of nonzero topological
sectors. A quick estimate indicates that the direct calculation of the
susceptibility up to $T=2$~GeV with reasonable accuracy would be approximately
ten orders of magnitude beyond present-day computational resources.

Instead of a straightforward calculation of the susceptibility, here we
propose to use the integral method. It is based on the simple fact that
derivatives of the weights of different topological sectors are much easier to
evaluate than the susceptibility itself. Indeed, a simple calculation reveals
that
\be
  \frac{d \log Z_Q/Z_0}{d \log T} = 
       \frac{d \beta}{d \log a} \langle S_g \rangle_{Q-0}
     + \sum_f \frac{d \log m_f}{d \log a} 
                 \langle \bar{\psi}_f \psi_f \rangle_{Q-0},
   \label{eq:derzq}
\ee
where $Z_Q/Z_0$ is the relative weight of topological sectors $Q$ and $0$ and
we introduced the notation 
\be
  \langle X \rangle_{Q-0} = \langle X \rangle_{Q} - \langle X \rangle_{0}  
\ee
for the difference of expectations of the quantity $X$ in the topological
sector $Q$ and zero. The right-hand side of Eq.~\ref{eq:derzq} can be easily
evaluated by measuring the gauge action $\langle S_g \rangle$ and the quark
condensates $\langle \bar{\psi}_f \psi_f \rangle $ separately in the charge $Q$
and charge zero sectors via fixed topological sector simulations. 

For the derivatives to be useful we also need the relative weight of different
topological sectors at some fixed temperature $T_0$. We can choose $T_0$ to be
low enough that a direct measurement of the weights is feasible.  Starting
from $T_0$ the weights can be integrated up to the desired higher temperature
where they can be combined to provide the susceptibility.

\begin{figure}[thb]
  \centering
  \includegraphics[width=0.8\textwidth,clip]{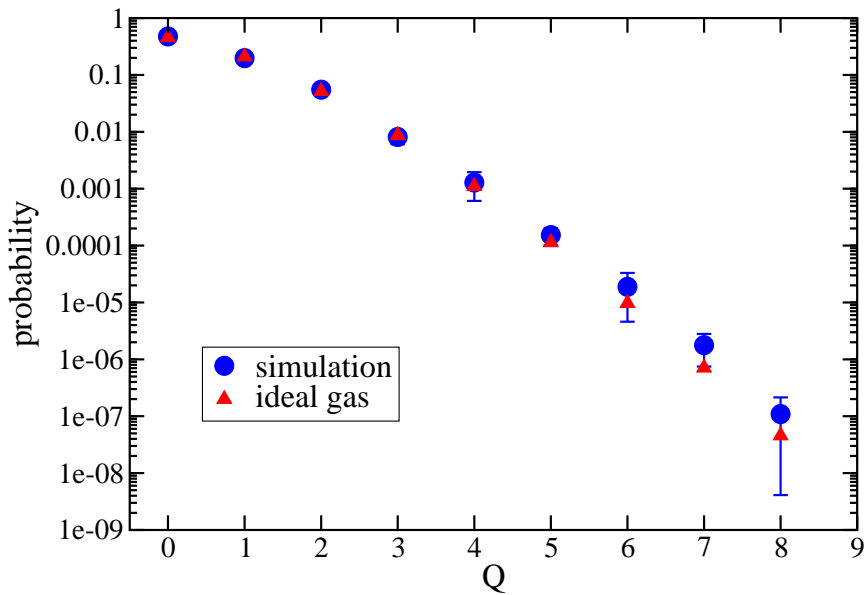}
  \caption{The probability of different topological sectors at $T=180$~MeV and
    $N_t=16$ in a system of spatial linear size $L=6.6$~fm. A comparison is
    made between our simulation data obtained with reweighting and a gas of
    free instantons with its density adjusted to provide the same topological
    susceptibility.}
  \label{fig:qweight}
\end{figure}

Even further simplification occurs since at high temperature the instanton gas
is so dilute that the weight of different sectors is precisely given by a free
instanton gas. If this is the case, the probability distribution of the
topological charge is already determined by $Z_1/Z_0$. Thus the susceptibility
can also be written in terms of this quantity and the weight of higher sectors
is not needed. In fact, already a bit above the cross-over a gas of free
instantons provides a rather accurate description of the system. To show this,
in Fig.~\ref{fig:qweight} we compare the charge probability distribution
obtained from our simulation with reweighting with that of a gas of free
instantons. The density of the free gas was chosen to reproduce the
topological susceptibility of the simulation data. We note that the system is
at a fairly low temperature, just above the cross-over and it has a large
spatial linear size $6.6$~fm potentially accommodating several instantons. In
spite of that we do not see a significant difference between the simulation
and the free gas. 

Up to $T=300$~MeV direct simulations are also available and compatible with
results obtained with the integral method. It is only above $300$~MeV that we
resort to only the integral method and the assumption of free instantons. At
that point the instanton gas is much more dilute than at the temperature shown
in the figure so the free gas should provide the appropriate
description. However, this only means that the interaction among instantons is
neglected but the one-instanton weight $Z_1/Z_0$ in the partition function is
still determined fully non-perturbatively. We emphasize that our approach is
different from what is known in the literature as the DIGA (dilute instanton
gas approximation) where the perturbative result is used for the one-instanton
weight.

\begin{figure}[thb]
  \centering
  \includegraphics[width=0.8\textwidth,clip]{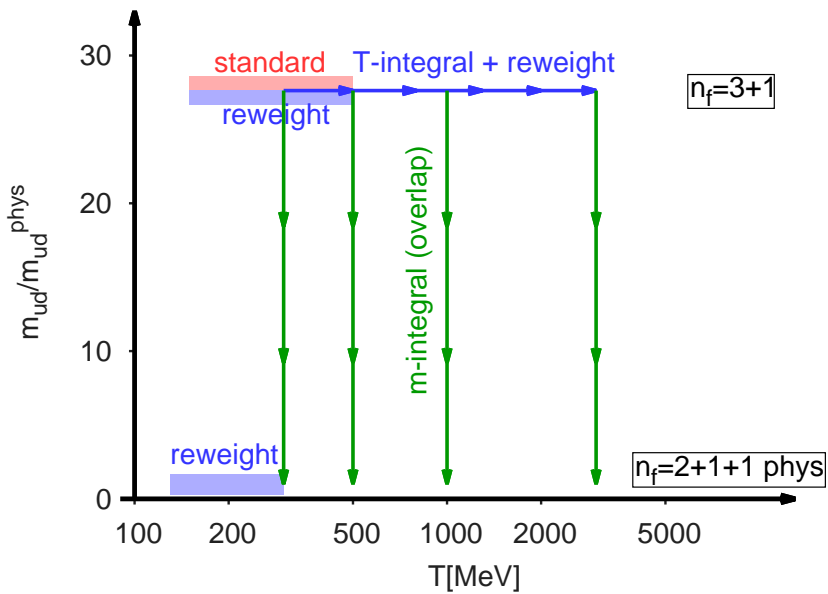}
  \caption{Schematic view of the simulations we have performed in the
    temperature versus light quark mass plane. The arrows indicate the paths
    of integration for the topological sector weights. We also marked the
    regions where only reweighting (no integration) and where the standard
    method (no reweighting, no integration) was used.}
  \label{fig:sims}
\end{figure}

Additional savings in the computing resources can be had by noticing that
integration in terms of the light quark masses is cheaper than integration in
the temperature. Therefore at the reference point $T_0$ we can use a system
with heavy $u$ and $d$ quarks that are chosen to be degenerate with the
physical $s$ quark. Due to the heavier $u$ and $d$ quarks the temperature
integration of the weights in this $N_f=3+1$ flavor system is much cheaper
than it would be in the physical $N_f=2+1+1$ system. To obtain the
susceptibility for physical quark masses, we can integrate the weights in
terms of the $u$ and $d$ quark masses starting from the $N_f=3+1$ theory at
the desired temperature. This integration in terms of the light quark masses
was carried out using $N_f=2+1$ overlap simulations. In Fig.~\ref{fig:sims} we
depict a schematic view of the simulations in the temperature versus $(u,d)$
quark mass plane. The arrows indicate the paths of integration for determining
the weights of topological sectors. We also show in the figure the regions
where different methods (reweighting without integration and the standard
method without reweighting or integration) were used independently. Throughout
the overlapping regions the different methods gave consistent results.

\section{Conclusions}

\begin{figure}[thb]
  \centering
  \includegraphics[width=0.8\textwidth,clip]{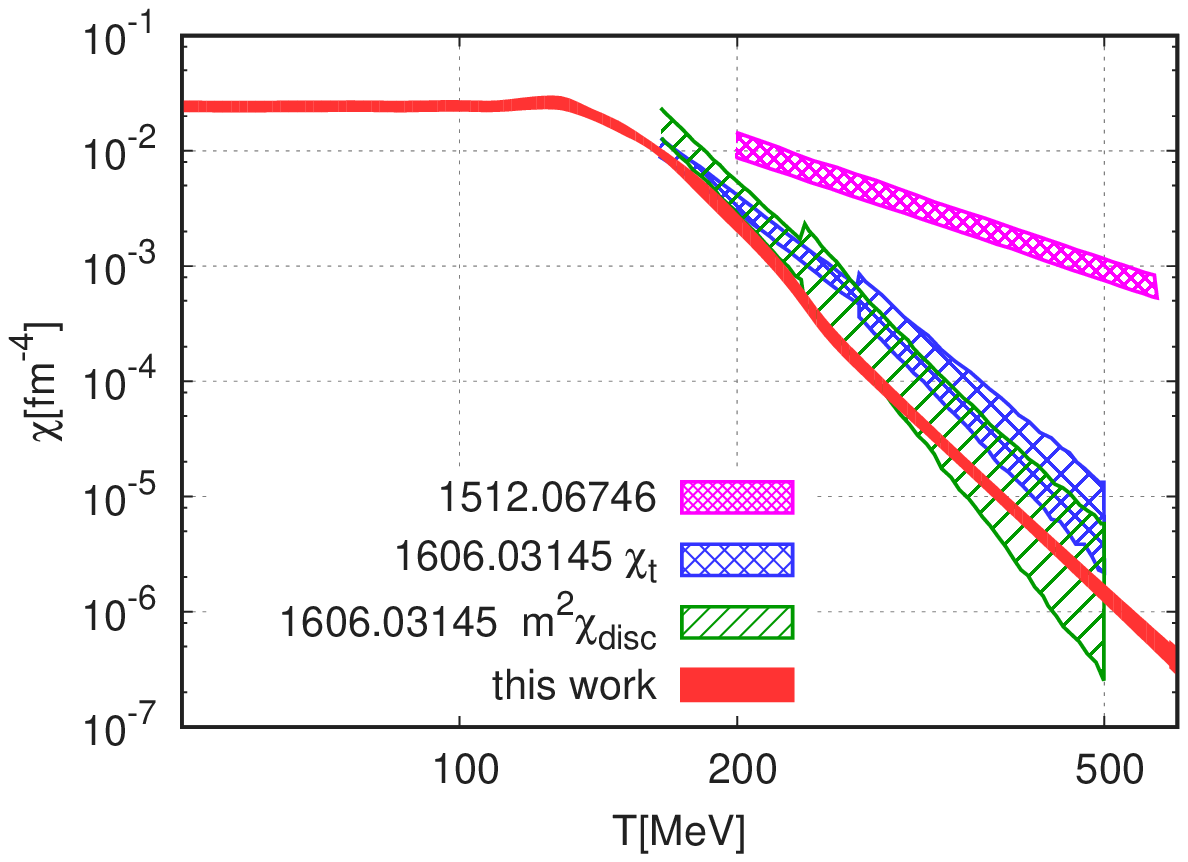}
  \caption{Our main result, the temperature-dependence of the QCD topological
    susceptibility. For comparison we also included two other results from the
    literature. 1512.06746 refers to Bonati et
    al. \cite{Bonati:2015vqz,Bonati:2017nhe} and 1606.03145 to Petreczky et
    al. \cite{Petreczky:2016vrs}. In each case the bands indicate the
    uncertainty quoted by the authors.}
  \label{fig:cmp}
\end{figure}

Finally, in Fig.~\ref{fig:cmp} we summarize our main result, the
temperature-dependence of the QCD topological susceptibility. In the same
figure we compare our results to two other determinations of the
susceptibility that appeared recently in the literature. Our results appear to
be significantly lower than those of Bonati et
al.\ \cite{Bonati:2015vqz}. However, the susceptibility obtained by Petreczky
at al.\ \cite{Petreczky:2016vrs} with one of their methods is consistent with
ours, albeit our results have much smaller uncertainties compared to theirs.
In addition, our results, going all the way up to $T=2$~GeV, cover a much
wider temperature range than the other results in the literature. 

The important point we would like to emphasize is that the topological
susceptibility at high temperature has much larger cutoff corrections than
most other quantities. Lattices that are sufficiently fine for typical QCD
thermodynamics are much too coarse to obtain a reliable continuum
extrapolation of the topological susceptibility. We demonstrated that these
unusually large lattice artifacts are due to small instantons and the
inability of non-chiral Dirac operators to properly describe their zero
modes. We showed that lattice zero modes occurring far away from zero and cannot
provide enough suppression for higher topological sectors and the topological
susceptibility can be grossly over-estimated. We also argued that this problem
can be solved by shifting the lattice zero modes to zero where their continuum
counterparts would occur and reweighting lattice configurations accordingly. An
alternative possibility would be to use a chiral quark action, like the
overlap where lattice zero modes are protected by an exact Ginsparg-Wilson
symmetry. 

\bibliography{axion.bib}
\end{document}